\documentclass[envcountsame,envcountsect]{llncs}

\usepackage[T1]{fontenc}
\usepackage{lmodern}
\usepackage{xcolor}
\usepackage[english]{babel}
\usepackage{algorithm}
\usepackage{algorithmic}
\usepackage[citestyle=numeric]{biblatex}

\usepackage{amsmath,amsfonts,amssymb}
\usepackage{graphics, graphicx}
\usepackage{calc}
\usepackage{url}
\usepackage{etoolbox}
\usepackage{options}
\RequirePackage{xparse}
\RequirePackage{xargs}
\RequirePackage{footnote}

\title{BSEA-1 - A Stream Cipher Backdooring Technique\thanks{This work has been presented at the Ruscrypto 2019 conference in Moscow}}  
\titlerunning{Mathematical Backdoors in Symmetric Encryption Systems}
\author{Eric Filiol}
\authorrunning{E. Filiol}
\tocauthor{Eric Filiol}
\institute{%
  Operational Cryptology and Virology Lab, ESIEA,\\
  38 rue des Drs Calmette et Gu\'erin, 53000 Laval, France,\\
  \email{\{filiol\}@esiea.fr}%
}

\begin{document}
\maketitle


\begin{abstract}
Recent years have shown that more than ever governments and intelligence agencies try to control and bypass the cryptographic means used for the protection of data. Backdooring encryption algorithms is considered as the best way to enforce cryptographic control. Implementation backdoors (at the protocol/implementation/ma\-nagement level) are generally considered. In this paper we propose to address the most critical issue of backdoors: mathematical backdoors or by-design backdoors, which are put directly at the mathematical design of the encryption algorithm. Considering a particular family (among all the possible ones) of backdoors, we present BSEA-1, a stream cipher algorithm which contains a design backdoor enabling an effective cryptanalysis. The BSEA-1 algorithm uses a 120-bit key. The exploitation of the backdoor enables to break the cipher with around 2 Kbits of knowplaintext in a few seconds. 
\end{abstract}

\section{Introduction}

Despite the fact that in the late 90s/early 2000s, citizens have partially obtained the freedom for using cryptography, the recent years have shown that more than ever, governments and intelligence agencies still try to control and bypass the cryptographic means used for the protection of data and of private life. Snowden's leaks were a first upheaval. A tremendous number of secret projects (from NSA, GCHQ) have been revealed to the public opinion which proves this situation. 

While the need for the security of everyday life activities (for companies, for citizens) requires more and more cryptography and highly secure communications means, recent bothering initiatives or decision by political decision-makers ask for an even stronger control over cryptography not to say preparing  the simple prohibition or ban of cryptographic application such as telegram. The most recent decision towards the control over cryptography is that of the Australian government \cite{bbc1} which makes mandatory the use of backdoor in encryption systems.  At the same time, the EU as well as a number of security agencies (such as French ANSSI, German BSI) confirmed that it is nonsense and militate for the mandatory use of end-to-end encryption.  

In this paper we address the most critical issue of backdoors: mathematical or by-design backdoors. In other words, the backdoor is put directly at the mathematical design of the encryption algorithm. The RSA's \texttt{Dual\_EC\_DRBG} standard case falls within this category \cite{Bernstein2016}. Other non-public examples are known within the military cryptanalysis community, and partially revealed to the public thanks to the 1995 Hans B\"uhler case \cite{strehle1994verschlusselt}. There is quite no public work on that topic. It is the technical realm of a few among the most eminent intelligence agencies (namely NSA, GCHQ) which moreover have the ability and power to step in and to influence the international standardization processes in one direction or another. Recently Bannier \& Filiol \cite{Filiol17} have published a block cipher algorithm (BEA-1) which is similar to the AES and which contains a mathematical backdoor enabling an effective cryptanalysis. This block cipher algorithm (80-bit block, 120-bit key size, 11 rounds) was designed to resist to linear and differential cryptanalyses.

This paper is organized as follows. In Section~\ref{s1} we discuss the comparative feasibility of backdoors in stream ciphers and block ciphers. We also present the state-of-the-art, history and previous work regarding backdoors. In Section~\ref{s2}, we describe our backdoored stream cipher algorithm BSEA-1 (standing for \textit{Backdoored Stream Encryption Algorithm 1}) and address the cryptographic security of this cipher, with respect to known cryptanalyses. In Section~\ref{s3} we explain how to exploit the backdoor when considering known plaintext and ciphertext only cryptanalyses. Finally in Section~\ref{conc} we summarizes our work and present future work.

\section{The Concept of Backdoor in Symmetric Cryptology} \label{s1}
The general concept of backdoor has been addressed in \cite{Filiol17}. In this section, we just deal with this concept when considering stream ciphers in comparison with block ciphers.
\subsection{Stream Ciphers vs Block Ciphers}
Their respective complexity is totally different, especially with respect to their combinatorial complexity. We can define the combinatorial complexity as the number of internal configurations that can be realized during the different steps of operation (key setup, encryption, decryption).
\begin{itemize}
\item \textit{Stream ciphers}. Their design complexity is rather low since they mostly rely on algebraic primitives (LFSRs and Boolean functions which have intensely been studied in the open literature). The cryptographic properties of these primitives are well-known may it be for the LFSRs or the Boolean functions used since the latter have generally a limited dimension (the number of input bits, 4 or 5 at most). Until the late 70s, backdoors relied on the fact that quite all algorithms were proprietary and hence secret. It was then easy to hide non primitive polynomials, weak combining Boolean functions or more exotic designs. The Hans B\"uhler case in 1995 \cite{strehle1994verschlusselt} shed light on that particular case. Being secret, anyone who use the algorithm may only perform a deep statistical analysis. In this respect, the pseudo-running key (which is combined to the plaintext or the ciphertext) must always exhibit excellent randomness properties.
\item \textit{Block ciphers}. This class of encryption algorithms is rather recent (end of the 70s for the public part). They exhibit so a huge combinatorial complexity that it is reasonable to think to backdoors. As described in \cite{daemen2002design} for a $k$-bit secret key and a $m$-bit input/output block cipher there are $((2^m)!)^{2^k}$ possible such block ciphers. For such an algorithm, the number of possible internal states (which involves both the key and the input block whereas stream ciphers just input a secret key) is so huge that we are condemned to have only a local view of the system, that is, the round function or the basic cryptographic primitives. We cannot be sure that there is no degeneration effect at a higher level. This  point has been addressed in \cite{daemen2002design} when considering linear cryptanalysis. Therefore, it seems reasonable to think that this combinatorial richness of block cipher may be used to hide backdoors.    
\end{itemize}
\subsection{Previous Work}
While a few research work does exist regarding backdoors in block ciphers (see \cite[Section 2]{Filiol17}), there is not public research work on stream ciphers, to the author's knowledge. It is somehow surprising when considering that from the mid-80s to the early 2000s, this class of encryption systems has been widely studied. At the industry level (that of encryption machines sold to governments), stream ciphers were also the vast majority of systems used throughout the world. Since, there are still used, at least partly, in payTV systems, telecommunications and satellite communication (where fast encryption is more than ever required), access control systems, subway tickets, and various other security-related applications \cite{nohl1,nohl2,nohl3}... With the rise of IoT, stream ciphers seem also to know some sort of come back.   

As far as the intelligence world is concerned, NSA and GCHQ \textemdash~among possibly a few others \textemdash~have conducted an intense research activity with regards to backdoors for this class of ciphers. The B\"uhler case in 1995 \cite{strehle1994verschlusselt} revealed that Crypto AG, a Swiss company and the main provider of cipher machines for nearly 120 governments and international entities, was working closely with the NSA to introduce backdoors in the encryption systems sold. So did a handful of other European companies selling crypto-machines.

Another kind of backdoors in stream ciphers relates to implementations that enables the secret key to be reused or to be the same with a high probability. For instance, the message key (used along with a base key) may have a very short entropy. In this case the cryptanalysis becomes rather easy \cite{fllbh10}.  
\section{Description of BSEA-1} \label{s2}
The BSEA-1 algorithm (standing for \textit{Backdoored Stream Encryption Algorithm version 1}) is based on our research work and the analysis of the rare available technical details and exchanges with experts in the field.

\begin{figure}
\begin{center}
	\includegraphics[width=\textwidth]{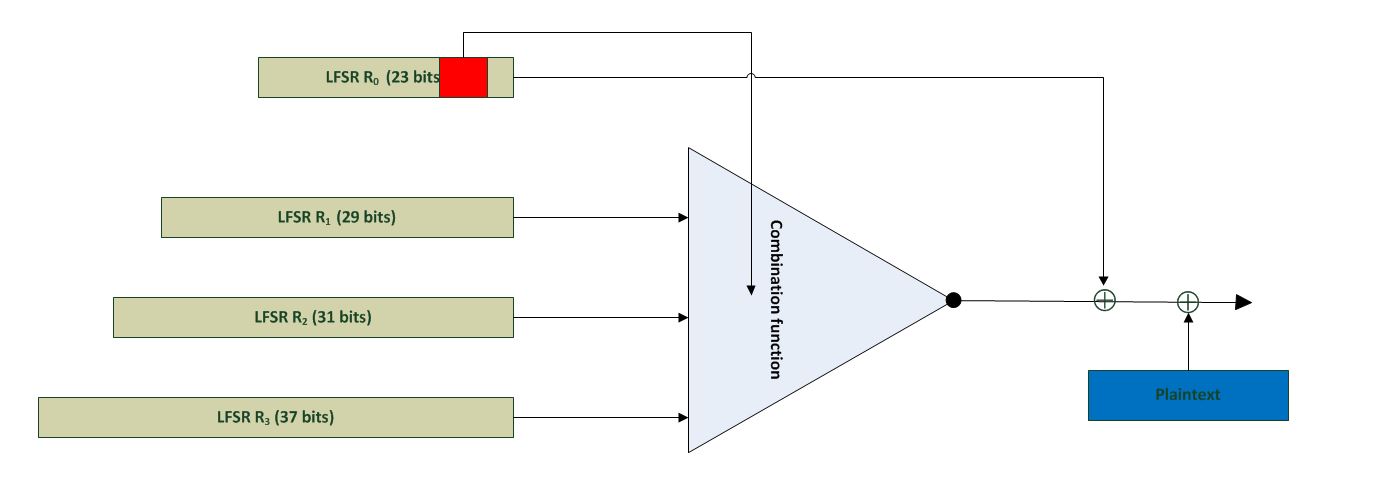}
\end{center}
\caption{General structure of BSEA-1 Algorithm} \label{bsea1}
\end{figure}

This section is intended to describe this cipher precisely. It is a classical combination generator \cite[Section 5.2]{Rueppel} which uses a 120-bit secret key (Figure~\ref{bsea1}). The essential difference with this design lies in the fact that the truth table of the combining Boolean function is modified at each time instant $t$ by one of the registers which is moreover irregularly clocked. 

The cryptographic primitives are the following:
\begin{itemize}
\item Four Linear Feedback Shift Registers $R_0, R_1, R_2$ and $R_3$ of respective length $L_0 = 23, L_1 = 29, L_2 = 31$ and $l_3 = 37$. Their initialization is the secret key at time instant $t = 0$.
\item The four feedback polynomials are primitive and given by
\begin{equation*}
\begin{split}
P_0(x) = &  x^{23} \oplus x^{22} \oplus x^{20} \oplus x^{18} \oplus x^{17} \oplus x^{13} \oplus x^{11}  \oplus x^{10} \oplus x^9 \oplus x^8 \oplus x^4 \oplus x^3+ \oplus x^2 \oplus x \oplus 1 \\
P_1(x) = &  x^{29} \oplus x^{28} \oplus x^{27} \oplus x^{25} \oplus x^{24} \oplus x^{23} \oplus x^{22} \oplus x^{21} \oplus x^{18} \oplus x^{17} \oplus x^{13} \oplus x^{11} \oplus x^{10} \oplus x^6 \\
&  \oplus x^5 \oplus x^3 \oplus x^2 \oplus x \oplus 1\\
P_2(x) = &  x^{31} \oplus x^{30} \oplus x^{27} \oplus x^{25} \oplus x^{24} \oplus x^{23} \oplus x^{22} \oplus x^{21} \oplus x^{20} \oplus x^{16} \oplus x^{15} \oplus x^{13} \oplus x^{12} \oplus x^{11} \\
&   \oplus x^{10} \oplus x^9 \oplus x^8 \oplus x^4 \oplus x^3 \oplus x \oplus 1 \\
P_3(x) = &  x^{37} \oplus x^{34} \oplus x^{33} \oplus x^{32} \oplus x^{30} \oplus x^{29} \oplus x^{26} \oplus x^{24} \oplus x^{20} \oplus x^{19} \oplus x^{18} \oplus x^{17} \oplus x^{16} \oplus x^{13}  \\
&  \oplus x^{11} \oplus x^8 \oplus x^7 \oplus x^6 \oplus x^4 \oplus x^2 \oplus 1 
\end{split}
\end{equation*}
\item The initial value of the Boolean function at time instant $t$ is given by 
\[ 0x6B = (0, 1, 1, 0, 1, 0, 1, 1) \]
\end{itemize}

A pseudo-random sequence $(\sigma_t)_{1 \leq t \leq N}$ is produced and xored to the plaintext (encryption) or to the ciphertext (decryption). The encryption algorithm is given hereafter:
\begin{figure*}
  \begin{algorithm}[H]
  	\floatname{algorithm}{\small{Algorithm}}
  	\small{
  		\caption{\small{Pseudo-random sequence generation (base version, encryption)}} \label{alg1}
  		\algsetup{
  			linenosize=\small,
  		}
  		\begin{algorithmic}[1]
  			\REQUIRE Secret 120-bit key $K$ and $P = (p_1, \ldots, p_N)$ a plaintext of length $N$ \\ \medskip
  			
  			\STATE Key setup $(R_0, R_1, R_2, R_3) \leftarrow K$
  			\STATE Combining function $f \leftarrow 0x6B$      \COMMENT{\textcolor{red}{$(1, 1, 0, 1, 0, 1, 1, 0)$}}
  			\FOR {$t$ from $1$ to $N$}
  			\STATE Compute $S = (R_0 \;\&\; 3) + 1$   \COMMENT{\textcolor{red}{Step value in $[1, 4]$}}
  			\FOR {$i$ from $1$ to $S$}
  			\STATE Clock register $R_0$ once
  			\ENDFOR
  			\STATE $X_0^t \leftarrow R_0 \; \& \; 1$    \COMMENT{\textcolor{red}{$R_0$ output $x_0^t$}}
  			\STATE $\tau \leftarrow (R_0 >> 3)\; \& \; 0x7$    
  			\STATE $f \leftarrow f \oplus (R_0 \;>>\; \tau)\; \&\; 0xFF$ \COMMENT{\textcolor{red}{modification pattern for $f$}}
  			\STATE Clock registers $R_1, R_2, R_3$ once and output $x_1^t, x_2^t, x^t_3$
  			\STATE $\sigma_t = f(x_1^t + (x_2^t << 1) + (x_3^t << 2)) \oplus x_0^t$
  			\ENDFOR 
  			\RETURN $(c_t = \sigma_t \oplus p_t)_{1 \leq t \leq N}$
  	\end{algorithmic}}
  \end{algorithm}	
\end{figure*}

The base algorithm presented here is the base version. A large number of variants can be derived from the base version:
\begin{itemize}
\item Feedback polynomials and Boolean function initial value can be changed.
\item Registers $R_1 R_2, R_3$ can be irregularly clocked by register $R_0$ according to different clocking settings.
\item Value $S$ (step value, line 4 in Algorithm~\ref{alg1}) and $\tau$ (line 9 in Algorithm~\ref{alg1}) can be computed according a lot of different ways.
\end{itemize}
\subsection{BSEA-1 Security Analysis}
In stream cipher theory, a number of cryptographic properties for the core primitives must be achieved:
\begin{itemize}
\item All feedback polynomials are primitive \cite{Rueppel}.
\item Feedback polynomial degrees are co-prime ($L_0 = 23, L_1 = 29, L_2 = 31, L_3 = 37)$ \cite{Rueppel}.
\item Each feedback polynomial has a prime degree (in order prevent the decimation attack \cite{filiol_indocrypt}).
\item Combination Boolean function (initial value) has relatively good cryptographic properties. 
\end{itemize}
The variability over the time provides a \textbf{false sense} of cryptographic security. Indeed, since the truth table is constantly changing, it \textbf{seems} impossible to build the
data required for known attacks:
\begin{itemize}
\item Noisy equations for correlation attacks \cite{sieg1} and fast correlation attacks \cite{Meier1989} and similar variants.
\item Non-linear equations to be solved in algebraic attacks and similar variants \cite{Bard}.
\end{itemize}
The statistical analysis of the pseudo-random sequence expanded from the secret key is also a very important cryptographic property. Since the design may be most of the time secret and embedded in a device (crypto-machine for instance or a IoT device), it is however possible to check the randomness properties in a black-box approach. In this respect BSEA-1 has been tested with different suites: FIPS 140-2/STS (US NIST standard), TestUI01 \cite{testui01} and DieHarder \cite{dieharder}. The final conclusion is that BSEA-1 is statistically compliant with FIPS 140-2. It means that BSEA-1 would then pass all classical cryptographic validation which are generally considered by the industry.
\section{BSEA-1 Cryptanalysis with the Backdoor Knowledge} \label{s3}
\subsection{Description of the Backdoor}
The value of the Boolean function varies over the time. So does its Walsh spectrum.
\begin{itemize}
	\item The Walsh transform summarizes the correlation between the Boolean function inputs and its output value \cite{meier2}
	\[\widehat{\chi_f}(u) = \sum_{x \in \mathbb{F}_2^n} -1^{f(x) \oplus <x, u>} \mbox{ and } P[f(x) = <x, u>] = \frac{1}{2}(1 + \frac{\widehat{\chi_f}(u)}{2^n}) \]
	\item The Walsh spectrum $\mathcal{S}$ gives the correlation for all the possible linear combination of the function inputs $u = (u_1, u_2, \ldots, u_n)$:
	\[  S = (\widehat{\chi_f}(00\cdots00), \widehat{\chi_f}(00\cdots01), \ldots, \widehat{\chi_f}(11\cdots 11)) \]
\end{itemize} 
Let us now see how to apply this on a Boolean function which is changing at each time instant
\begin{itemize}
	\item Whenever the Boolean function takes particular values, the Walsh spectrum takes strong correlation values (backdoor values)
	      For instance when $f = 0x69$ then $S = (0, 0, 0, 0, 0, 0, 0, -8)$
	\item For these particular values, it means that the linear combination of the inputs and the output are equal with probability $p = 1.0$. It is then possible to write a linear equation whose unknowns are the $R_1, R_2$ and $R_3$ key bits.
	
	\item Exactly 16 values over 256 possibles have a similar Walsh spectrum.
		\begin{equation*}
		\begin{split}
		\mathcal{B} = & \{0x69, 0x5A, 0x55, 0x3C, 0x33, 0xF, 0xF0, 0xCC, 0xC3, 0xAA, 0xA5, 0x99, 0x99, 0x96, \\
		&  0x66, 0x00, 0xFF\}
		\end{split}
		\end{equation*}
		\item Values $0x00$ and $0xFF$ enables to speed up the cryptanalysis by keeping or discarding key candidates quickly.
\end{itemize}
It is worth mentioning that the time indices for the backdoor values are not the same from key to key. They are strongly dependent from the secret key. 
\subsection{Known Plaintext Attack}
In this case we know $N$ bits of ciphertext ($(c_t)_{1 \leq t \leq N}$) and of corresponding plaintext ($(p_t)_{1 \leq t \leq N}$). Hence we can obtain the pseudo-random sequence in a straightforward way: $\sigma_t = c_t \oplus p_t$ for each time instant $t$. The cryptanalysis considers an exhaustive search with respect to register $R_0$. Here are the main steps:
\begin{itemize}
	\item Exhaustive search on register $R_0$. \medskip
	\item For each initial value $I_0$ of $R_0$ 
	\begin{itemize}
		\item Boolean function values $0x00$ and $0xFF$ enables to discard $I_0$. 
		\item We build a system of 97 equations of 97 unknowns and solve it.
		\item We test the final $K$ (23 + 97 bits) against the known plaintext.
	\end{itemize}
\end{itemize}

The pseudo-code of the attack is given in Algorithm~\ref{alg2}.
\begin{algorithm}[H]
\floatname{algorithm}{\small{Algorithm}}
\small{
		\caption{\small{BSEA-1 Cryptanalysis Algorithm (Known Plaintext Attack)}} \label{alg2}
		\algsetup{
			linenosize=\small,
		}
		\begin{algorithmic}[1]
			\REQUIRE Pseudo-random sequence $(\sigma_t)_{0 \leq t \leq N}$ \\ \medskip
			
			\FOR {$I_0$ from $0$ to $2^{L_0} - 1$}
			\STATE $R_0 \leftarrow I_0$
			\FOR {$t$ from $1$ to $N$}
			\STATE Compute Boolean function value $f_{I_0^t}$
			\IF {($f_{I_0^t} == 0x00 \mbox{ and } \sigma_t \not = 0$) or ($f_{I_0^t} == 0xFF \mbox{ and } \sigma_t \not = 1)$}
			\STATE Discard $I_0$ and continue
			\ENDIF
			\IF {$f_{I_0^t} \in \mathcal{B} \setminus \{0x00, 0xFF \}$}
			\STATE Write Equation and add it to the system $S_{I_0}$
			\ENDIF
			\ENDFOR
			\STATE Solve equation system $S_{I_0}$
			\STATE Test the solution $K_{I_0}$ against $(\sigma_t)_{0 \leq t \leq N}$
			\IF {$K_{I_0}$ is correct}
			\STATE $K = K_{I_0}$ and break
			\ENDIF
			\ENDFOR
			
			\RETURN $K$
	\end{algorithmic}}
\end{algorithm}	
We need only $N = 1,800$ bits of known plaintext to break the whole key $K$ with a complexity of $\mathcal{O}(2^{L_0})$ where $L_0$ is the length of register $R_0$. 
In most real-life cases, side (implementation) backdoors enable to have a few kbits of known plaintext very easily. For instance, in strategic telegraph systems
(export version), synchronizing information is encrypted. In synchronous systems (high data rate) an initial, agreed patterns of bits (generally defined in the technical specifications) is sent but encrypted. In asynchronous system (low data rate) synchronizing bits are inserted (called stop-bits and start-bits) and encrypted. 

When evaluating a
cryptographic system not only the algorithm but also its implementation and use must be carefully analysed in order to evaluate the risk of side implementation backdoors. Cryptographic algorithms must be armoured door installed on a cardboard wall.

\subsection{Ciphertext-only Attack}
The principle remains the same but requires a little bit more effort. Whenever the Boolean function takes particular values (different from those in $\mathcal{B}$), the Walsh spectrum  takes strong correlation values
\begin{itemize}
	\item For instance when $f = 0x7$ then $S = (2, -2, -2, 2, -6, -2, -2, 2)$ (exactly 128 values over 256 possibles have a similar Walsh spectrum).
	\item Then we have Equation $1 + x_3^t = f(x_3^t, x_2^t, x_1^t)$ holding with probability $p = 0.875$. 
	\item If we consider that $p[m_t = 0] = 0.6$ (probability of a plaintext bit to be equal to 0, which is a rather commonly observed for many languages and for most encodings) then Equation $1 + x_3^t = c_t$ holds with probability $p = 0.575$
\end{itemize}
We get then a system of noisy equations to solve for each register \cite{sieg1}. We need about $50\;kb$ of ciphertext bits to recover the whole key with a complexity of at most $\mathcal{O}(2^{54})$. However many optimizations are possible to reduce this complexity significantly.
\section{Conclusion and Future Work} \label{conc}
In this paper, we have proposed one possible technique of stream cipher backdooring at the design level. It is illustrated by a 120-bit algorithm, named BSEA-1 which exhibits many of the desirable properties that any secure stream cipher algorithm should. When exploiting the backdoor, we manage to break it with a very limited amount of resources successfully. 

The next BSEA variant consists in slightly changing the way the Boolean function is updated over the time. Instead of modifying the whole function, it is better to modify it ``by half''. The modification pattern $\pi$ of size $2^{n - 1}$ is then applied as follows:
\[ f \leftarrow f \oplus ((\pi << 2^{(n - 1)}) | \pi) \]
The cryptanalysis method becomes less obvious than for BSEA-1 and requires to consider far different cryptanalysis approaches and methods.

It is worth stressing on the fact that backdooring stream ciphers requires to consider working at the combination module mostly.
\begin{itemize}
	\item Secure primitives of the random engine part (LFSRs) are very much well known (primitive, dense polynomials of prime and coprime length...) 
	\item However due to lack of combinatorial complexity, backdoored designs are bound to remain secret mostly.
\end{itemize}
The next step in this research work about cryptographic backdooring techniques (design and detection) will be to consider more sophisticated designs and primitives such as Nonlinear Feedback Shift Registers (NLFSR), design with memory... 

\printbibliography

\end{document}